# TWO-DIMENSIONAL TOPOLOGICAL GRAVITY AND EQUIVARIANT COHOMOLOGY

E. GETZLER

The analogy between topological string theory and equivariant cohomology for differentiable actions of the circle group on manifolds has been widely remarked on. One of our aims in this paper is to make this analogy precise. We show that topological string theory is the "derived functor" of semi-relative cohomology, just as equivariant cohomology is the derived functor of basic cohomology. That homological algebra finds a place in the study of topological string theory should not surprise the reader, granted that topological string theory is the conformal field theorist's algebraic topology.

In [7], we have shown that the cohomology of a topological conformal field theory carries the structure of a Batalin-Vilkovisky algebra (actually, two commuting such structures, corresponding to the two chiral sectors of the theory). In the second part of this paper, we describe the analogous algebraic structure on the equivariant cohomology of a topological conformal field theory: we call this structure a gravity algebra. This algebraic structure is a certain generalization of a Lie algebra, and is distinguished by the fact that it has an infinite sequence of independent operations $\{a_1, \ldots, a_k\}$, $k \geq 2$, satisfying quadratic relations generalizing the Jacobi rule. (The operad underlying the category of gravity algebras has been studied independently by Ginzburg-Kapranov [9].)

The author is grateful to M. Bershadsky, E. Frenkel, M. Kapranov, G. Moore, R. Plesser and G. Zuckerman for the many ways in which they helped in the writing of this paper; also to the Department of Mathematics at Yale University for its hospitality while part of this paper was written.

## 1. Equivariant cohomology

Equivariant cohomology for differentiable circle actions was developed by A. Borel and H. Cartan around 1950. (See Atiyah-Bott [2] for references.) It is a contravariant functor from the category of manifolds with differentiable circle actions to graded vector spaces $M \mapsto H^{\bullet}_{S^1}(M)$, characterized by the following properties.

(1) (Normalization) If the circle action on $M$ is free, then $H^{\bullet}_{S^1}(M) \cong H^{\bullet}(M/S^1)$.

The author is partially supported by a fellowship of the Sloan Foundation and a research grant of the NSF.

(2) (Homotopy invariance) If $f : M_1 \to M_2$ is an equivariant map inducing a homotopy equivalence, then $f : H^\bullet_{S^1}(M_2) \to H^\bullet_{S^1}(M_1)$ is an isomorphism.

(3) (Meyer-Vietoris) If $M = U \cup V$, where $U$ and $V$ are invariant open submanifolds of $M$, there is a long exact sequence

$$\ldots \to H^{\bullet-1}_{S^1}(U \cap V) \to H^\bullet_{S^1}(M) \to H^\bullet_{S^1}(U) \oplus H^\bullet_{S^1}(V) \to H^\bullet_{S^1}(U \cap V) \to \ldots$$

The example of the sphere

$$S^\infty = \{(z_0, z_1, \ldots) \in \mathbb{C}^\infty \mid |z_0|^2 + |z_1|^2 + \cdots = 1\}$$

is fundamental. The circle acts freely, by sending $(z_0, z_1, \ldots)$ to $(e^{i\theta} z_0, e^{i\theta} z_1, \ldots)$, and has quotient $\mathbb{CP}^\infty$. But $S^\infty$ is contractible, showing that the equivariant cohomology of a point

$$H^\bullet_{S^1}(*) \cong H^\bullet(\mathbb{CP}^\infty) \cong \mathbb{C}[\Omega]$$

is isomorphic to a polynomial algebra in a variable $\Omega$ of degree 2.

Homotopy invariance shows that if $V$ is any linear representation of the circle group, then

$$H^\bullet_{S^1}(V) \cong H^\bullet_{S^1}(*) \cong \mathbb{C}[\Omega].$$

As another example, consider the rotation action of the circle on the two-sphere $S^2$. Using the standard cover of $S^2$ by $U = S^2 \setminus \{\infty\}$ and $V = S^2 \setminus \{0\}$, we see that the Meyer-Vietoris long exact sequence splits into exact sequences

$$0 \to H^0_{S^1}(S^2) \to H^0_{S^1}(U) \oplus H^0_{S^1}(V) \to H^0_{S^1}(U \cap V) \to H^1_{S^1}(U \cap V) \to 0$$

and for $i \geq 2$, $H^i_{S^1}(S^2) \cong H^i_{S^1}(U) \oplus H^i_{S^1}(V)$. It follows that

$$H^\bullet_{S^1}(S^2) \cong \{f(\Omega) \oplus g(\Omega) \in \mathbb{C}[\Omega] \mid f(0) = g(0)\}.$$

There is a de Rham-type construction of equivariant cohomology, due to H. Cartan. If $T$ is the vector field generating the circle action, let

$$\iota(T) : \Omega^\bullet(M) \to \Omega^{\bullet-1}(M)$$

be contraction by $T$, and let $\mathcal{L}(T) = [d, \iota(T)]$ be the Lie derivative with respect to $T$. Let $\Omega^\bullet(M)_{\text{basic}}$ be the basic subcomplex of $\Omega^\bullet(M)$, on which $\mathcal{L}(T)$ and $\iota(T)$ vanish. One might naively imagine that the equivariant cohomology of $M$ is the cohomology of the basic complex $\Omega^\bullet(M)$. However, this is incorrect, since the resulting theory does not satisfy the Meyer-Vietoris axiom. Indeed, taking the basic subcomplex is not an exact functor: it does not preserve exact sequences.

This is precisely the situation for which homological algebra was developed. One replaces the basic subcomplex by its "derived" analogue, which is the complex

$$W(\mathfrak{h}) \otimes \Omega^\bullet(M).$$

Here, $W(\mathfrak{h})$ is the Weil complex of the Lie algebra $\mathfrak{h}$ of the circle group. Explicitly, it is a free graded commutative algebra on two generators $\omega$ and $\Omega$, of degree 1 and 2 respectively, with differential $\delta$ the unique derivation such that $\delta\omega = \Omega$ and $\delta\Omega = 0$. The Weil complex is contractible, with contracting homotopy

$$h\Omega^n = n\omega\Omega^{n-1} \quad \text{and} \quad h\omega\Omega^n = 0.$$

Thus, the inclusion $(\Omega^\bullet(M), d) \to (W(\mathfrak{h}) \otimes \Omega^\bullet(M), \delta + d)$ of complexes induces an isomorphism of cohomology.

Let $\iota$ be the unique derivation on $W(\mathfrak{h})$ such that $\iota\omega = 1$ and $\iota\Omega = 0$. Then the operator $\iota + \iota(T)$ on $W(\mathfrak{h}) \otimes \Omega^\bullet(M)$ is exact; the effect is just as if we had replaced $M$ by $S^\infty \times M$, on which the circle acts freely. This shows that the equivariant cohomology of $M$ is the cohomology of the basic subcomplex of $W(\mathfrak{h}) \otimes \Omega^\bullet(M)$, on which $\iota + \iota(T)$ and $\mathcal{L}(T)$ vanish.

Let us make the basic subcomplex more explicit. The kernel of $\iota + \iota(T)$ equals

$$\{\alpha(\Omega) - \omega(\iota(T)\alpha(\Omega)) \mid \alpha \in \Omega^\bullet(M)[\Omega]\}.$$

The differential of such an element equals

$$(\delta + d)(\alpha(\Omega) - \omega(\iota(T)\alpha(\Omega))) = (1 - \omega\iota(T))(d - \Omega\iota(T))\alpha(\Omega) + \omega\mathcal{L}(T)\alpha(\Omega).$$

Thus, the basic subcomplex may be identified with the space of polynomials in $\Omega$, with coefficients in

$$\Omega^\bullet(M)(0) = \{\alpha \in \Omega^\bullet(M) \mid \mathcal{L}(T)\alpha = 0\},$$

and with differential $d - \Omega\iota(T)$. This is H. Cartan's complex.

Observe that the equivariant cohomology may be defined on any complex $\mathcal{E}$ with operators $\iota : \mathcal{E}^\bullet \to \mathcal{E}^{\bullet-1}$ and $\mathcal{L} : \mathcal{E}^\bullet \to \mathcal{E}^\bullet$ such that $[d, \iota] = \mathcal{L}$ and $\iota^2 = 0$. The invariant subcomplex $\mathcal{E}(0)$ is the kernel of $\mathcal{L}$, the basic subcomplex $\mathcal{E}_{\text{basic}}$ is the intersection of the kernels of $\iota$ and $\mathcal{L}$, while the equivariant cohomology $H^\bullet_{S^1}(\mathcal{E})$ of $\mathcal{E}$ is the cohomology of the differential $d - \Omega\iota$ on $\mathcal{E}(0)[\Omega]$. The following lemma shows that the cohomology of the Cartan complex satisfies the normalization axiom of equivariant cohomology.

**Lemma 1.1.** *If the action of $\iota$ is exact and the action of $\mathcal{L}$ is semisimple ($\mathcal{E}$ splits into a sum of eigenspaces of $\mathcal{L}$), then the inclusion of complexes*

$$(\mathcal{E}_{\text{basic}}, d) \hookrightarrow (\mathcal{E}(0)[\Omega], d - \Omega\iota)$$

*induces an isomorphism on cohomology.*

*Proof.* For $k > 0$, let $\Omega^{-k}$ be the operator on $\mathcal{E}(0)[\Omega]$

$$\Omega^{-k}(a^n + \Omega a^{n+2} + \Omega^2 a^{n+4} + \dots) = (a^{n+2k} + \Omega a^{n+2(k+1)} + \Omega^2 a^{n+2(k+2)} + \dots).$$

Since $\iota$ is exact, there is an operator $h : \mathcal{E}^\bullet \to \mathcal{E}^{\bullet+1}$ such that $[\iota, h] = 1$; since the action of $\mathcal{L}$ is semisimple, we may suppose that $h$ commutes with $\mathcal{L}$, and hence preserves $\mathcal{E}(0)$. For $k \geq 0$, let $H_k$ be the operator on $\mathcal{E}(0)[\Omega]$

$$H_k = \Omega^{-k-1} h[d, h]^k.$$

Since $[\iota, [d, h]] = 0$, we see that

$$[d - \Omega\iota, H_k] = \Omega^{-k-1}[d, h]^{k+1} - \Omega^{-k}[d, h]^k - (1 - \Omega\Omega^{-1})\Omega^{-k}\iota h[d, h]^k,$$

where of course $1 - \Omega\Omega^{-1}$ is the projection from $\mathcal{E}(0)[\Omega]$ onto $\mathcal{E}(0)$ given by extracting the constant term.

The operator $H = \sum_{k=0}^{\infty} (-1)^k H_k$ is well defined, since for any $v \in \mathcal{E}$, $\Omega^{-k} v = 0$ for $k$ sufficiently large. It follows that

$$[d - \Omega\iota, H] = (1 - \Omega\Omega^{-1}) \sum_{k=0}^{\infty} (-1)^k \Omega^{-k} \iota h [d, h]^k - 1,$$

and thus that $H$ defines a homotopy contraction to the image of $(1 - \Omega\Omega^{-1})\iota h$, which is nothing other than the basic subcomplex $\mathcal{E}_{\text{basic}}$. $\square$

For any $\mathbb{C}[\Omega]$-module $W$, there is a variant $H^\bullet_{S^1}(M, W)$ of equivariant cohomology, the cohomology of the complex $\Omega^\bullet(M) \otimes W$ with differential $d - \Omega\iota(T)$. This satisfies a universal coefficient sequence

$$0 \to H^\bullet_{S^1}(M) \otimes_{\mathbb{C}[\Omega]} W \to H^\bullet_{S^1}(M; W) \to \text{Tor}^{\mathbb{C}[\Omega]}(H^{\bullet-1}_{S^1}(M), W) \to 0.$$

An important example is the periodic equivariant cohomology $\Omega^{-1} H^\bullet_{S^1}(M)$, defined by taking as coefficients the $\mathbb{C}[\Omega]$-module $\mathbb{C}((\Omega^{-1}))$ of Laurent series in $\Omega^{-1}$; this theory may be thought of as the "torsion-free" part of $H^\bullet_{S^1}(M)$. Another example is the theory $G^\bullet_{S^1}(M)$ of Jones [10], with coefficients $\mathbb{C}((\Omega^{-1}))/\Omega\mathbb{C}[\Omega]$. There is a long exact sequence

$$\ldots \to G^{i-1}_{S^1}(M) \to H^{i-2}_{S^1}(M) \to \Omega^{-1} H^i_{S^1}(M) \to G^i_{S^1}(M) \to \ldots$$

induced by the short-exact sequence of $\mathbb{C}[\Omega]$-modules

$$0 \to \Omega\mathbb{C}[\Omega] \to \mathbb{C}((\Omega^{-1})) \to \mathbb{C}((\Omega^{-1}))/\Omega\mathbb{C}[\Omega] \to 0.$$

This sequence may be compared to the Bockstein long exact sequence

$$\ldots \to H^{i-1}(X, \mathbb{Q}/\mathbb{Z}) \to H^i(X, \mathbb{Z}) \to H^i(X, \mathbb{Q}) \to H^i(X, \mathbb{Q}/\mathbb{Z}) \to \ldots$$

of ordinary cohomology, induced by the short-exact sequence of $\mathbb{Z}$-modules (abelian groups) $0 \to \mathbb{Z} \to \mathbb{Q} \to \mathbb{Q}/\mathbb{Z} \to 0$.

2. THE SEMI-INFINITE WEIL COMPLEX OF THE VIRASORO ALGEBRA

Recall the definition of the $N = 2$ superconformal Virasoro algebra. This is the one-parameter family of graded Lie algebras $\mathfrak{g}[s]$, where $s \in \mathbb{R}$, with bosonic fields $L(z)$ and $J(z)$ of spin 2 and 1, and fermionic fields $G^{\pm}(z)$ of spin $\frac{3}{2} \mp s$, satisfying the operator product expansions

$$L(z) \cdot L(w) \sim (\tfrac{3}{2} - 2s^2)\frac{d}{(z-w)^4} + \frac{2L(w)}{(z-w)^2} + \frac{\partial L(w)}{z-w},$$

$$L(z) \cdot G^{\pm}(w) \sim (\tfrac{3}{2} \mp s)\frac{G^{\pm}(w)\,dw}{(z-w)^2} + \frac{\partial G^{\pm}(w)\,dw}{z-w},$$

$$L(z) \cdot J(w) \sim -\frac{2sd}{(z-w)^3} + \frac{J(w)}{(z-w)^2} + \frac{\partial J(w)}{z-w},$$

$$G^{+}(z) \cdot G^{-}(w) \sim \frac{d}{(z-w)^3} + \frac{J(w)}{(z-w)^2} + \frac{L(w) + (\tfrac{1}{2} - s)\partial J(w)}{z-w},$$

$$J(z) \cdot J(w) \sim \frac{d}{(z-w)^2},$$

$$J(z) \cdot G^{\pm}(w) \sim \pm\frac{G^{\pm}(w)}{z-w}, \qquad G^{\pm}(z) \cdot G^{\pm}(w) \sim 0.$$

Here, $d$ is a real number, called the central charge. Note that the fields $L(z)$ and $J(z)$ generate respectively a Virasoro algebra and a $U(1)$ current algebra. When we write $(z-w)^{-1}$ in the operator product expansion, we are following the conventions of the physicists: more properly, $(z-w)^{-1}$ should be understood to represent the one-form $\frac{dw}{z-w}$.

If $A(z)$ and $B(z)$ are fields of conformal dimension $a$ and $b$ respectively, with operator product

$$A(z) \cdot B(w) \sim \sum_k \frac{(A(k)B)(w)}{(z-w)^k},$$

where $(A(k)B)(z)$ is a field of conformal dimension $a+b-k$, the graded commutators of the Fourier modes of $A$ and $B$ are given by the formula

$$[A_n, B_m] = \sum_{k \geq 1} \binom{n+a-1}{k-1}(A(k)B)_{n+m}.$$

Expanding the fields of the $N = 2$ superconformal algebra in Fourier modes

$$L(z) = \sum_{n=-\infty}^{\infty} L_n z^{-n}\Big(\frac{dz}{z}\Big)^2, \qquad J(z) = \sum_{n=-\infty}^{\infty} J_n z^{-n}\frac{dz}{z},$$

$$G^{\pm}(z) = \sum_{n=-\infty}^{\infty} G_n^{\pm} z^{-n}\Big(\frac{dz}{z}\Big)^{\frac{3}{2}\mp s},$$

we obtain the explicit graded commutation relations

$$[L_n, L_m] = (n - m)L_{n+m} + (\tfrac{1}{4} - 2s^2)dn(n^2 - 1)\delta^0_{n+m},$$
$$[L_n, G^\pm_m] = ((\tfrac{1}{2} \mp s)n - m)G^\pm_{n+m},$$
$$[L_n, J_m] = -mJ_{n+m} - sdn(n+1)\delta_{n+m},$$
$$[G^+_n, G^-_m] = L_{n+m} + ((\tfrac{1}{2} + s)n - (\tfrac{1}{2} - s)m)J_{n+m} + \tfrac{1}{2}d((n-s)^2 - \tfrac{1}{4})\delta_{n+m},$$
$$[J_n, J_m] = dn\delta_{n+m},$$
$$[J_n, G^\pm_m] = \pm G^\pm_{n+m}, \qquad [G^\pm_n, G^\pm_m] = 0.$$

For $k \in \mathbb{Z}$, there is an isomorphism between $\mathfrak{g}[s]$ and $\mathfrak{g}[s + k]$, the twisting map of Witten and Eguchi-Yang, which replaces $L(z)$ by $L(z) - k\partial J(z)$. This has the following effect on the modes:

$$\tau^k L_n = L_n + k(n+1)J_n, \qquad \tau^k G^\pm_n = G^\pm_{n \pm k}, \qquad \tau^k J_n = J_n.$$

A similar map, with $k \in \mathbb{Z} + \tfrac{1}{2}$ odd, defines an isomorphism between $\mathfrak{g}[s]$ and a Neveu-Schwarz analogue of $\mathfrak{g}[s + k]$.

We will be interested in the case $s = \tfrac{1}{2}$, the topological Virasoro algebra. (Note that this is the same as the untwisted Neveu-Schwartz $N = 2$ Virasoro algebra $\mathfrak{g}[0]$.) In this case, the currents $G^\pm(z)$ have spin 1 and 2, and the Virasoro current $L(z)$ is anomaly-free. The topological Virasoro algebra $\mathfrak{g}[\tfrac{1}{2}]$ is isomorphic to $\mathfrak{g}[-\tfrac{1}{2}]$, by the replacement of the field $L(z)$ by $L(z) + \partial J(z)$.

For the purposes of this paper, a topological Virasoro module is a graded module $\mathcal{E}$ for the topological Virasoro algebra which satisfies three additional conditions.

(1) The eigenvalues of the operators $J_0$ and $L_0$ are integers.
(2) The eigenspace $(L_0, J_0) = (h, q)$ is finite dimensional for all $h$ and $q$.
(3) For any vector $v \in \mathcal{E}$, the subspace of $\mathcal{E}$ obtained by acting on $v$ by the universal enveloping algebra of $\mathbf{n} = \text{span}\{L_n, G^\pm_n, J_n \mid n > 0\}$ is finite-dimensional.

These hypotheses allow us among other things to manipulate infinite sums meaningfully; they are the analogues for the topological Virasoro algebra of the category $\mathcal{O}$ of a simple Lie algebra introduced by Bernstein-Gelfand-Gelfand.

A topological Virasoro module $\mathcal{E}$ is, in particular, a complex, graded by the operator $J_0$, with differential $G^+_0$, the BRST operator. This chain complex $\mathcal{E}$ carries the same structure as the differential forms on a manifold with circle action: the analogue of the Lie derivative is $L_0$, while the analogue of $\iota(T)$ is $G^-_0$. The $\mathfrak{h}$-basic subcomplex of $\mathcal{E}$ is the subspace

$$\mathcal{E}_{\text{basic}} = \{v \in \mathcal{E} \mid L_0 v = G^-_0 v = 0\},$$

while the invariant subcomplex is $\mathcal{E}(0) = \{v \in \mathcal{E} \mid L_0 v = 0\}$. We will prove that the space of physical states of the associated topological string theory may be identified with the equivariant cohomology $H^\bullet_{S^1}(\mathcal{E})$ of $\mathcal{E}$, with differential $G_0^+ - \Omega G_0^-$, where $\Omega$ may be identifed with the zero mode of the bosonic ghost $\gamma$.

We can now define the analogue of the Weil complex in the setting of topological Virasoro modules, studied by Feigin and Frenkel [5] under the name of the semi-infinite Weil complex. Introduce the fermionic oscillators $\{b_n, c^n \mid n \in \mathbb{Z}\}$, with commutation relations $[b_n, c^m] = \delta_n^m$, and bosonic oscillators $\{\beta_n, \gamma^n \mid n \in \mathbb{Z}\}$, with commutation relations $[\beta_n, \gamma^m] = \delta_n^m$.

For each integer $p \in \mathbb{Z}$, we define the semi-infinite Weil complex $W[p]$ to be the representation of this algebra with vacuum $|p\rangle$ characterized by

$$b_n|p\rangle = \beta_n|p\rangle = 0, \quad n \geq p,$$
$$c^n|p\rangle = \gamma^n|p\rangle = 0, \quad n < p.$$

The integer $p$ is called the picture of the semi-infinite Weil complex. These representations are inequivalent for different values of $p$. (Note that our $p$ corresponds to $1-p$ in the notation of Feigin-Frenkel.) To be more precise, $W[p]$ is the tensor product of the free graded commutative algebra in generators $\{c^n, \gamma^n \mid n \geq p\}$ with the dual of the free graded commutative algebra in generators $\{c^n, \gamma^n \mid n < p\}$.

The fields of the semi-infinite Weil complex have the mode expansions

$$c(z) = \sum_{n=-\infty}^{\infty} c^n z^n \left(\tfrac{dz}{z}\right)^{-1}, \quad b(z) = \sum_{n=-\infty}^{\infty} b_n z^{-n} \left(\tfrac{dz}{z}\right)^{2},$$
$$\gamma(z) = \sum_{n=-\infty}^{\infty} \gamma^n z^n \left(\tfrac{dz}{z}\right)^{-1}, \quad \beta(z) = \sum_{n=-\infty}^{\infty} \beta_n z^{-n} \left(\tfrac{dz}{z}\right)^{2},$$

with operator products

$$b(w) \cdot c(z) \sim \beta(z) \cdot \gamma(w) \sim \frac{1}{z-w},$$

while $c(z) \cdot c(w) \sim b(z) \cdot b(w) \sim \gamma(z) \cdot \gamma(w) \sim \beta(z) \cdot \beta(w) \sim 0$.

From these operator products, it follows that normal ordering of a monomial $:\ldots:$ in generators $\{b_n, c^n, \beta_n, \gamma^n\}$ is defined by moving all operators $b_n$ and $\beta_n$ to the left if $n < -1$ and to the right if $n \geq -1$, paying attention to the sign convention in exchanging two fermionic operators. These odd conventions come about because the mode $b_n$ of a free field of spin 2 corresponds to the mode $b_{n+1}$ of a free field of spin 1. Alternatively, we could choose the usual normal ordering convention, but modify the operator products $b(z) \cdot c(w)$ and $\beta(z) \cdot \gamma(w)$, as in Feigin-Frenkel [5].

**Theorem 2.1.** *The semi-infinite Weil complex $W[p]$ is a topological Virasoro module*

*with central charge $-3$, with action given by the fields*

$$\mathbb{L}(z) = 2\partial c(z)b(z) + c(z)\partial b(z) + 2\partial\gamma(z)\beta(z) + \gamma(z)\partial\beta(z),$$
$$\mathbb{G}^+(z) = \gamma(z)b(z),$$
$$\mathbb{G}^-(z) = b(z) + 2\partial c(z)\beta(z) + c(z)\partial\beta(z),$$
$$\mathbb{J}(z) = c(z)b(z) + 2\gamma(z)\beta(z).$$

*The zero-modes act on the vacuum $|p\rangle$ by the formulas $\mathbb{J}_0|p\rangle = -(p+1)|p\rangle$ and $\mathbb{L}_0|p\rangle = \mathbb{G}_0^\pm|p\rangle = 0$.*

The theorem may be proved either by means of the operator product expansion, or directly by expanding the fields in modes:

$$\mathbb{L}_n = \sum_i (i-n)( {:}c^i b_{i+n}{:} + {:}\gamma^i \beta_{i+n}{:} ),$$
$$\mathbb{G}_n^+ = \sum_i \gamma^i b_{i+n},$$
$$\mathbb{G}_n^- = b_n + \sum_i (i-n) c^i \beta_{i+n},$$
$$\mathbb{J}_n = \sum_i ( {:}c^i b_{i+n}{:} + 2{:}\gamma^i \beta_{i+n}{:} ).$$

Since the operators are all quadratic in the oscillators, it suffices to check that the Poisson brackets of the associated classical observables are as they should be (which we leave to the reader), and then check the commutation relations on the vacuum $|p\rangle$. We now use the fact that $\mathbb{J}_n|p\rangle = 0$ for $n \geq 0$, while

$$\mathbb{J}_{-n}|p\rangle = \sum_{i=p}^{p+n-1} (c^i b_{i-n} + 2\gamma^i \beta_{i-n}).$$

Similar formulas hold for the other operators. It follows that for $n > 0$,

$$[\mathbb{J}_n, \mathbb{J}_{-n}]|p\rangle = \sum_{i=p}^{p+n-1} (c^{i-n} b_i c^i b_{i-n} + 4\gamma^{i-n}\beta_i \gamma^i \beta_{i-n})|p\rangle$$
$$= \sum_{i=p}^{p+n-1} (1-4)|p\rangle = -3n|p\rangle.$$

Similarly,

$$[\mathbb{G}_n^+, \mathbb{G}_{-n}^-]|p\rangle = \sum_{i=p}^{p+n-1} (i-2n) c^{i-n}\beta_i \gamma^i b_{i-n}|p\rangle$$
$$= \sum_{i=p}^{p+n-1} (i-2n)|p\rangle = \left(n(p+1) - \tfrac{3n(n+1)}{2}\right)|p\rangle.$$

The commutators $[\mathbb{L}_n, \mathbb{J}_{-n}]|p\rangle$ and $[\mathbb{L}_n, \mathbb{L}_{-n}]|p\rangle$ are calculated in the same way. In particular, the Virasoro algebra spanned by the operators $\mathbb{L}_n$ has vanishing central

charge, since the contributions of the $b - c$ and $\beta - \gamma$ fields cancel: the first has $c = -26$, while the second has $c = 26$.

If $\mathcal{E}$ is an auxilliary topological Virasoro module $\mathcal{E}$ with central charge $d$ (called topological matter), the tensor product $W[p] \otimes \mathcal{E}$ is a topological Virasoro module with central charge $d - 3$, with action given by the sum of the two actions on $W[p]$ and $\mathcal{E}$. Following Eguchi et al. [4], this structure of a topological Virasoro module on $W[p] \otimes \mathcal{E}$ will be referred to as the matter picture; the total fields will be denoted $\mathbb{L}_{\mathrm{tot}}(Z)$, $\mathbb{G}_{\mathrm{tot}}^{\pm}(z)$ and $\mathbb{J}_{\mathrm{tot}}(z)$.

Denote by $K$ the operator on $W[p] \otimes \mathcal{E}$, introduced by Eguchi et al. [4],
$$K = \mathrm{Res}(c(z)\partial c(z)\beta(z) + c(z)G^-(z)) = \sum_{i<j}(i-j)c^i c^j \beta_{i+j} + \sum_i c^i G_i^-.$$

Here, Res denotes the residue at $z = 0$, while $G^-(z)$ is the field which represents the action of the modes $G_n^-$ of $\mathfrak{g}[\tfrac{1}{2}]$ on $\mathcal{E}$. The analogue of this formula for the Weil complex of a finite-dimensional Lie algebra appears to be folklore among topologists, although we have found no explicit reference.

By the assumptions on the module $\mathcal{E}$, the operator $K$ is locally finite: that is, given a vector $|v\rangle \in W[p] \otimes \mathcal{E}$, $K^i|v\rangle = 0$ for $i \gg 0$. The exponential $U = \exp(K)$ is well-defined, commutes with $\mathbb{L}_{\mathrm{tot}}(z)$ and $\mathbb{J}_{\mathrm{tot}}(z)$, and
$$\begin{aligned} U\mathbb{G}_{\mathrm{tot}}^+(z)U^{-1} &= c(z)\partial c(z)b(z) + c(z)(2\partial\gamma(z)\beta(z) + \gamma(z)\partial\beta(z) + L(z)) \\ &\quad + \gamma(z)(b(z) - G^-(z)) - \partial(c(z)\mathbb{J}_{\mathrm{tot}}(z)) + \tfrac{d-3}{2}\partial^2 c(z) + G^+(z), \\ U^{-1}\mathbb{G}_{\mathrm{tot}}^-(z)U &= b(z). \end{aligned}$$

We call this structure of a topological Virasoro module on $W[p] \otimes \mathcal{E}$ the string picture, since the differential $U(\mathbb{G}_{\mathrm{tot}}^+)_0 U^{-1}$ is the usual BRST differential of the semi-infinite cohomology in the tensor product of $\mathcal{E}$ and the Fock space of the fields $\beta(z)$ and $\gamma(z)$ (the total derivative terms proportional to $\partial(c(z)\mathbb{J}_{\mathrm{tot}}(z))$ and $\partial^2 c(z)$ do not contribute to the zero-mode, since the field $U\mathbb{G}_{\mathrm{tot}}^+ U^{-1}$ has conformal dimension 1). The existence of the operator $U$ shows that the matter and string pictures are isomorphic. The formula
$$K|p\rangle = \sum_{p \leq i < p/2} \sum_{i<j<p-i} (i-j)c^i c^j \beta_{i+j}|p\rangle,$$
shows that the operator $U$ leaves the vacuum $|p\rangle$ invariant if $p \geq -1$.

We will now calculate the space of physical states of topological string theory, that is, the cohomology of the basic sub-complex $(W[p] \otimes \mathcal{E})_{\mathrm{basic}}$ with differential $(\mathbb{G}_{\mathrm{tot}}^+)_0$.

**Theorem 2.2.** *There are natural identifications*
$$H^\bullet((W[p] \otimes \mathcal{E})_{\mathrm{basic}}, (\mathbb{G}_{\mathrm{tot}}^+)_0) \cong \begin{cases} H_{S^1}^\bullet(\mathcal{E}), & p \leq 0, \\ G_{S^1}^\bullet(\mathcal{E}), & p > 0. \end{cases}$$

*Proof.* For $i \geq p$, the operator $[\mathbb{G}_0^+, c^i \beta_i] = c^i b_i + \gamma^i \beta_i$ acting on $W[p]$ has positive spectrum, with kernel the set of vectors $|v\rangle$ such that $b_i|v\rangle = \beta_i|v\rangle = 0$, while for $i < p$, it has negative spectrum, with kernel the set of vectors $|v\rangle$ such that $c^i|v\rangle = \gamma^i|v\rangle = 0$. Let $\chi$ be the function

$$\chi(i) = \begin{cases} +1, & i \geq p \text{ and } i \neq 0, \\ -1, & i < p \text{ and } i \neq 0, \\ 0, & i = 0. \end{cases}$$

The operator

$$H = \sum_{i \neq 0} \chi(i) c^i \beta_i$$

commutes with $\mathbb{L}_0$, $\mathbb{G}_0^-$, $L_0$ and $G_0^\pm$, and $[\mathbb{G}_0^+, H]$ is positive; denote by $\mathcal{V}$ its kernel. Thus, $H$ furnishes a chain homotopy between $(W[p] \otimes \mathcal{E})_{\text{basic}}$ and $\mathcal{V}_{\text{basic}}$. The vector space $\mathcal{V}$ may be described explicitly as follows:

- ($p \leq 0$) $\mathcal{V}$ may be thought of as the Weil complex of the circle tensored with $\mathcal{E}$, by associating to $\Omega^n$ the vector $(\gamma^0)^n|p\rangle$, and to $\omega\Omega^n$ the vector $c^0(\gamma^0)^n|p\rangle$.
- ($p > 0$) $\mathcal{V}$ may be thought of as the dual of the Weil complex of the circle tensored with $\mathcal{E}$, by associating to $\Omega^{-n}$ the vector $b_0 \beta_0^n|p\rangle$, and to $\omega\Omega^{-n}$ the vector $\beta_0^n|p\rangle$.

Let us concentrate on the case $p \leq 0$. The cohomology of $\mathcal{V}_{\text{basic}}$ when $p > 0$ may be calculated in much the same way: it equals $G^\bullet_{S^1}(\mathcal{E})$. The differential $(\mathbb{G}_{\text{tot}}^+)_0$ on $\mathcal{V}$ is given by the formula

$$(\mathbb{G}_{\text{tot}}^+)_0 (f(\gamma^0) + c^0 g(\gamma^0)) = (G_0^+ f(\gamma^0) + c^0 g(\gamma^0)) - c^0 G_0^+ g(\gamma^0),$$

where $f, g \in \mathcal{E}[\gamma^0]$ are polynomials in $\gamma^0$. The basic subcomplex of $\mathcal{V}$ equals

$$\mathcal{V}_{\text{basic}} = \{f(\gamma^0) - c^0 G_0^- f(\gamma^0) \mid L_0 f(\gamma^0) = 0\},$$

and has differential

$$(\mathbb{G}_{\text{tot}}^+)_0 (f(\gamma^0) - c^0 G_0^- f(\gamma^0)) = (G_0^+ - \gamma^0 G_0^-) f(\gamma^0).$$

This is just the complex $(\mathcal{E}(0)[\gamma^0], G_0^+ - \gamma^0 G_0^-)$ defining $H^\bullet_{S^1}(\mathcal{E})$. □

## 3. Application to topological string theory

Let $\mathcal{E}$ be the space of states of a topological conformal field theory. Owing to the presence of both chiral and anti-chiral sectors, $\mathcal{E}$ is acted on by two commuting topological Virasoro algebras $\{L(z), G^\pm(z), J(z)\}$ and $\{\bar{L}(\bar{z}), \bar{G}^\pm(\bar{z}), \bar{J}(\bar{z})\}$, each with its own semi-infinite Weil complex, $W[p]$ and $\bar{W}[p]$. Define the basic subcomplex of $W[p] \otimes \bar{W}[p] \otimes \mathcal{E}$ to be the intersection of the kernels of $L_0 - \bar{L}_0$ and $G_0^- - \bar{G}_0^-$. The results of the last section extend immediately to this setting, proving that for $p \leq 0$, the cohomology of the basic subcomplex $(W[p] \otimes \bar{W}[p] \otimes \mathcal{E})_{\text{basic}}$ is the cohomology of the complex
$$\Big(\mathcal{E}(0)[\Omega], (G_0^+ + \bar{G}_0^+) - \Omega(G_0^- - \bar{G}_0^-)\Big),$$
where $\mathcal{E}(0) = \{v \in \mathcal{E} \mid L_0 v = \bar{L}_0 v\}$. The operator $\Omega$ may be identified with $\frac{1}{2}(\gamma_0 - \bar{\gamma}_0)$, as is seen by the formula
$$\gamma_0 G_0^- + \bar{\gamma}_0 \bar{G}_0^- = \tfrac{1}{2}(\gamma_0 + \bar{\gamma}_0)(G_0^- + \bar{G}_0^-) + \tfrac{1}{2}(\gamma_0 - \bar{\gamma}_0)(G_0^- - \bar{G}_0^-).$$

If $\mathcal{E}$ is the space of states of a topological conformal field theory, there is a vector $|0\rangle \in \mathcal{E}$, the vacuum, which is invariant under the action of the Lie algebra $\mathbf{sl}(2, \mathbb{C})$ spanned by $\{L_1, L_0, L_{-1}, \bar{L}_1, \bar{L}_0, \bar{L}_{-1}\}$; this vector represents the functional integral over a disk. The tensor product $W[p] \otimes \bar{W}[p] \otimes \mathcal{E}$ then has vacuum $|p\rangle \otimes |\bar{p}\rangle \otimes |0\rangle$; in the physical picture, this vacuum must be invariant under $\mathbf{sl}(2, \mathbb{C})$, in other words, the vector $|p\rangle$ should be preserved by the operators $\{\mathbb{L}_1, \mathbb{L}_0, \mathbb{L}_{-1}\}$. This singles out the $(-1)$-picture as the physical one, by the following lemma.

**Lemma 3.1.** $p = -1$ *is the unique value of $p$ for which $|p\rangle$ is annihilated by the Lie algebra $\mathbf{sl}(2)$ generated by $\{\mathbb{L}_1, \mathbb{L}_0, \mathbb{L}_{-1}\}$ acting on $W[p]$.*

*Proof.* For any picture, we have $\mathbb{L}_1 |p\rangle = \mathbb{L}_0 |p\rangle = 0$, while in the expression
$$\mathbb{L}_{-1}|p\rangle = -\sum_i (i+1)\Big(c^i b_{i-1} + \gamma^i \beta_{i-1}\Big)|p\rangle$$
the terms with $i = 1$ vanish if and only if $p = -1$. $\square$

For more on the physical significance of pictures, see Friedan-Martinec-Shenker [6].

The space of physical states of topological string theory in the string picture is the cohomology of the complex of vectors $|v\rangle \in W[-1] \otimes \bar{W}[-1] \otimes \mathcal{E}$ such that
$$(\mathbb{L}_{\text{tot}} - \bar{\mathbb{L}}_{\text{tot}})_0 |v\rangle = (b_0 - \bar{b}_0)|v\rangle = 0,$$
with differential $U^{-1}(\mathbb{G}_{\text{tot}}^+)_0 U$. We may also describe topological string theory in the matter picture, by the complex
$$(W[-1] \otimes \bar{W}[-1] \otimes \mathcal{E})_{\text{basic}} = \{|v\rangle \mid (\mathbb{L}_{\text{tot}} - \bar{\mathbb{L}}_{\text{tot}})_0 |v\rangle = (\mathbb{G}_{\text{tot}}^- - \bar{\mathbb{G}}_{\text{tot}}^-)_0 |v\rangle = 0\},$$
with differential $(\mathbb{G}_{\text{tot}}^+ + \bar{\mathbb{G}}_{\text{tot}}^+)_0$. As emphasized by Dijkgraaf, and by Eguchi et al. [4], the non-triviality of the interaction between gravity and matter in a topological string theory is reflected in the matter picture by the condition $(\mathbb{G}_{\text{tot}}^- - \bar{\mathbb{G}}_{\text{tot}}^-)_0 |v\rangle = 0$ defining the complex $(W[-1] \otimes \bar{W}[-1] \otimes \mathcal{E})_{\text{basic}}$.

In summary, we have the following definition.

**Definition 3.2.** The state space of a topological string theory is the semi-infinite Weil complex $W[-1] \otimes \bar{W}[-1] \otimes \mathcal{E}$, where $\mathcal{E}$ is the space of states of a topological conformal field theory. The space of physical states of the topological string theory is the cohomology of the basic complex $(W[-1] \otimes \bar{W}[-1] \otimes \mathcal{E})_{\text{basic}}$.

We see from the last section that there is a natural isomorphism between the space of physical states of the topological string theory and $H^\bullet_{S^1}(\mathcal{E})$. In particular, Lemma 1.1 shows that if the action of $G_0^- - \bar{G}_0^-$ on $\mathcal{E}$ has vanishing cohomology, then $H^\bullet_{S^1}(\mathcal{E})$ is naturally isomorphic to the cohomology of the total BRST differential $G_0^+ + \bar{G}_0^+$ on $\mathcal{E}_{\text{basic}}$. In the context of topological string theory, this is known as the semi-relative cohomology. In the language of homological algebra, we might say that "topological string theory is the derived functor of semi-relative cohomology." This explains the difficulty which theoretical physicists have had in pinning down a definition of topological string theory: in homological algebra, one resolution is as good as another.

For a topological conformal field theory obtained by twisting a *unitary* $N = 2$ superconformal field theory, the operators $G_0^+$ and $G_0^-$ vanish on $\mathcal{E}(0)$. Indeed, $\mathcal{E}$ has an inner product with respect to which $G_n^+ = (G_{-n}^-)^*$, and thus for $v \in \mathcal{E}$,

$$(v, L_0 v) = |G_0^+ v|^2 + |G_0^- v|^2.$$

This is consistent with Witten's result (Section 4 of [11]), that for the associated topological string theory, there is an isomorphism between the space of physical states and $\mathcal{H}[\Omega]$, where $\mathcal{H}$ is the cohomology of $\mathcal{E}$ with respect to $G_0^+$: his descendents $\sigma_k(\phi)$, where $\phi \in \mathcal{H}$ and $k \geq 0$, may be identified with our homology classes $\Omega^k \phi$.

A very different class of topological Virasoro modules may be obtained by a construction of Bershadsky-Lerche-Nemeschansky-Warner [3]. The following reformulation of their construction owes much to conversations with G. Moore.

The Virasoro algebra consists of vector fields on the circle. It is equally natural to consider the Lie algebra of first-order differential operators on the circle: to the differential operator $a(z)\partial + b(z)$ is associated the operator given by the residue $\text{Res}(a(z)l(z) + b(z)j(z))$, where $l(z)$ and $j(z)$ are fields of conformal dimension 2 and 1. Let $\mathcal{F}$ be a projective module over the Atiyah algebra for which the fields $l(z)$ and $j(z)$ have operator products

$$l(z) \cdot l(w) \sim \frac{1}{(z-w)^4} + \frac{2l(w)}{(z-w)^2} + \frac{\partial l(w)}{z-w},$$

$$l(z) \cdot j(w) \sim \frac{d+1}{(z-w)^3} + \frac{j(w)}{(z-w)^2} + \frac{\partial j(w)}{z-w},$$

$$j(z) \cdot j(w) \sim \frac{d-1}{(z-w)^2}.$$

**Theorem 3.3.** *The tensor product $\mathcal{E}$ of $\mathcal{F}$ with the fermionic Fock space generated by fields $b(z)$ and $c(z)$ carries an action of the twisted topological Virasoro $\mathfrak{g}[-\frac{1}{2}]$ with* *central charge d, given by fields*

$$L(z) = \partial c(z)b(z) + l(z),$$
$$G^+(z) = c(z)\partial c(z)b(z) + c(z)l(z) - \partial c(z)j(z) + \tfrac{d}{2}\partial^2 c(z),$$
$$G^-(z) = b(z),$$
$$J(z) = c(z)b(z) + j(z).$$

On performing the twist which replaces $L(z)$ by $L(z) + \partial J(z)$, we obtain a topological Virasoro module, whose differential

$$G_0^+ = \mathrm{Res}(c(z)\partial c(z)b(z) + c(z)(l(z) + \partial j(z)))$$

is the BRST operator associated to the Virasoro module with Virasoro action $l(z) + \partial j(z)$; this Virasoro module has central charge $c = 26$. The action of the operator $\mathbb{G}_0^-$ is exact (by the formula $[\mathbb{G}_0^-, c_0] = 1$), so the equivariant cohomology of this topological Virasoro module is naturally isomorphic to its basic cohomology, in other words the semi-relative cohomology.

An example is furnished by $D = 2$ string theory, in which we have two scalar fields $\phi_1(z)$ and $\phi_2(z)$, with

$$l(z) = -\tfrac{1}{2}\partial\phi_1(z)\partial\phi_1(z) - \tfrac{1}{2}\partial\phi_2(z)\partial\phi_2(z) + \sqrt{2}(\partial^2\phi_1(z) + i\partial^2\phi_2(z))$$

and $j(z) = -i\sqrt{2}\partial\phi_2(z)$. In this example, $d = 3$, so the associated topological Virasoro module $W[-1] \otimes \mathcal{E}$ has vanishing central charge.

## 4. Gravity algebras

In this section, we study the algebraic structure induced on the equivariant cohomology of a topological conformal field theory by the functional integrals of the theory. This algebraic structure is not the same as that induced by the correlation functions of topological gravity, in which the functional integrals are integrated over the fundamental class of the compactified moduli space $\overline{\mathcal{M}}_{0,n}$; our algebraic structure is obtained by integrating over cycles supported in the interior $\mathcal{M}_{0,n}$. However, the work of Ginzburg-Kapranov [9] reveals a clear relationship between these two algebraic structures, analogous to the duality between the Lie and commutative operads, which we expect will lead to a clearer understanding of the algebraic structures underlying topological gravity and string field theory.

Whereas in the previous sections, we have graded our complexes in the cohomological fashion, in this section we reverse conventions. To transfer between the homological and cohomological gradings of a complex, one simply applies the rule $V_i = V^{-i}$.

The operad which governs topological string theory may be most easily constructed in the context of operads in the symmetric monoidal category of mixed complexes.

**Definition 4.1.** A mixed complex is a graded vector space $V_\bullet$, together with operators $\delta : V_\bullet \to V_{\bullet-1}$ and $\Delta : V_\bullet \to V_{\bullet+1}$, such that $\delta^2 = \delta\Delta + \Delta\delta = \Delta^2 = 0$.

Mixed complexes model the homotopy theory of topological spaces with a circle action: the singular chain complex of such a space is a module over the Hopf algebra $H_\bullet(S^1)$. In particular, the homology of a space with circle action is a mixed complex, with vanishing $\delta$. We will use the notation $\Sigma V$ for the suspension of a graded vector space, for which $(\Sigma V)_i = V_{i-1}$.

The category of mixed complexes is a symmetric monoidal category, with tensor product
$$(V \otimes W, \delta \otimes 1 + 1 \otimes \delta, \Delta \otimes 1 + 1 \otimes \Delta),$$
The $k$-th tensor power of a mixed complex $V$ is denoted $V^{(k)}$. (All tensor products in this article will be graded tensor products: for example,
$$\Delta(v \otimes w) = \Delta v \otimes w + (-1)^{|v|} v \otimes \Delta w.)$$

An operad on the category of mixed complexes is a sequence $\mathbf{b}(k)$, $k \geq 0$, of mixed complexes on which the symmetric group $\mathbb{S}_k$ acts, together with the structure of a triple on the functor
$$V \mapsto \mathbb{T}(\mathbf{b}, V) = \sum_{k=0}^\infty \mathbf{b}(k) \otimes_{\mathbb{S}_k} V^{(k)}$$
from the category of mixed complexes to itself. An algebra over an operad $\mathbf{b}$ is a mixed complex $A$, together with maps of mixed complexes
$$\rho(k) : \mathbf{b}(k) \otimes_{\mathbb{S}_k} A^{(k)} \to A$$
inducing a structure map $\rho : \mathbb{T}(\mathbf{b}, A) \to A$ which makes of $A$ an algebra over $\mathbb{T}(\mathbf{b})$. For a more detailed discussion of operads and their algebras, see Getzler-Jones [8].

Let $D$ denote the closed unit disk of $\mathbb{C}$. Let $\mathcal{C}(k)$ be the space of all maps
$$(d(1), \ldots, d(k)) : \coprod_{i=1}^k D \to D$$
such that each map $d(i) : D \to D$ is affine, that is, of the form $z \mapsto az + b$ with $a$ real and $b$ complex, and such that the maps $d(i)$ have disjoint images. It is clear that the symmetric group $\mathbb{S}_k$ acts freely on $\mathcal{C}(k)$. The circle group $S^1$ acts on $\mathcal{C}(k)$ through its conjugation action on the affine group, which send $az + b$ to $az + e^{i\theta}b$.

If $\mathcal{C}$ is an operad in the category of spaces with circle action, we obtain an operad in the category of mixed complexes by setting
$$\mathbf{b}(k) = H_\bullet(\mathcal{C}(k)).$$
(This is a mixed complex whose differential vanishes.) If $\mathcal{C}$ is the little discs operad, the resulting operad in the category of mixed complexes is called the Batalin-Vilkovisky operad, and denoted $\mathbf{b}$. This operad has as its underlying linear operad the braid operad **b** of [7], and thus differs from the Batalin-Vilkovisky operad **bv** of that reference, which was an operad in the category of chain complexes, and not of mixed complexes. However, the categories of algebras of these two Batalin-Vilkovisky operads are identical. The following theorem is proved in Getzler [7].

**Theorem 4.2.** *A Batalin-Vilkovisky algebra is a mixed complex with the structure of a graded commutative algebra, such that $\delta$ is a derivation while*

$$\Delta(abc) = \Delta(ab)c + (-1)^{|a|}a\Delta(bc) + (-1)^{(|a|-1)|b|}b\Delta(ac)$$
$$- (\Delta a)bc - (-1)^{|a|}a(\Delta b)c - (-1)^{|a|+|b|}ab(\Delta c).$$

The equivariant homology of a space $M$ with free circle action may be identified with the kernel of the operator $\Delta$ acting on its ordinary homology, or in other words with the suspension $\Sigma H_\bullet(M/S^1)$ of the ordinary homology of its quotient by the circle action. This is consistent with the universal coefficient theorem for equivariant cohomology:

$$0 \to \Sigma \operatorname{Ext}^1_{\mathbb{C}[\Omega]}(H^{S^1}_\bullet(M), \mathbb{C}[\Omega]) \to H^\bullet_{S^1}(M) \to \operatorname{Ext}^0_{\mathbb{C}[\Omega]}(H^{S^1}_\bullet(M), \mathbb{C}[\Omega]) \to 0$$

Indeed, when the circle action action on $M$ is free, the equivariant homology $H^{S^1}_\bullet(M)$ is a torsion-module, so that

$$\operatorname{Ext}^i_{\mathbb{C}[\Omega]}(H^{S^1}_\bullet(M), \mathbb{C}[\Omega]) \cong \begin{cases} 0, & i = 0, \\ H^{S^1}_\bullet(M), & i = 1. \end{cases}$$

If **b** is an operad in the category of mixed complexes, then

$$\mathbf{m} = \ker(\Delta : \mathbf{b} \to \mathbf{b})$$

is a linear operad. In the special case that $\mathbf{b}(k) = H_\bullet(\mathcal{C}(k))$, where $\mathcal{C}$ is an equivariant operad such that the action of the circle on $\mathcal{C}(k)$ is free, we see that

$$\mathbf{m}(k) \cong H^{S^1}_\bullet(\mathcal{C}(k)) \cong \Sigma H_\bullet(\mathcal{C}(k)/S^1).$$

Now, in the special case that $\mathcal{C}$ is the little discs operad, the quotient of $\mathcal{C}(k)$ by the circle action is homotopy equivalent to the moduli space $\mathcal{M}_{0,k+1}$ of genus zero curves with $k+1$ marked points, so we see that

$$\mathbf{m}(k) \cong \Sigma H_\bullet(\mathcal{M}_{0,k+1}).$$

We will call the operad **m** the gravity operad; an algebra over it will be called a gravity algebra.

The following result makes explicit the action of $\Delta$ on **b**.

**Proposition 4.3.** *The dual $\mathbf{b}(k)^*$ of the graded vector space $\mathbf{b}(k)$ is the graded commutative ring with generators $\omega_{ij}$, $1 \leq i \neq j \leq k$, of degree 1, and relations*

(1) $\omega_{ji} = \omega_{ij}$, *and*

(2) $\omega_{ij}\omega_{jk} + \omega_{jk}\omega_{ki} + \omega_{ki}\omega_{ij} = 0$.

*The symmetric group $\mathbb{S}_k$ acts on $\mathbf{b}(k)^*$ by $\pi \cdot \omega_{ij} = \omega_{\pi(i)\pi(j)}$. The operator $\Delta$ is the adjoint of the derivation which sends the generator $\omega_{ij}$ to 1.*

*Proof.* All of this except the action of $\Delta$ is due to Arnold [1], and is proved by a simple induction on $k$. The action of $\Delta$ is easily understood by observing that the maps $\pi_{ij}$ are $S^1$-equivariant, and that $\Delta$ sends $\omega_{12} \in \mathbf{bv}(2)^*$ to $1 \in \mathbf{bv}(2)^*$. $\square$

The homology class in $\mathbf{b}(2)$ dual to 1 represents the product in a Batalin-Vilkovisky algebra, while the class dual to $\omega_{12}$ represents the bracket

$$(-1)^{|a_1|}[a_1, a_2] = \Delta(a_1 a_2) - (\Delta a_1)a_2 - (-1)^{|a_1|}a_1(\Delta a_2).$$

It is amusing to check explicitly that the action of $\Delta$ is compatible with the relations in $\mathbf{b}(k)^*$:

$$\Delta(\omega_{ij}\omega_{jk} + \omega_{jk}\omega_{ki} + \omega_{ki}\omega_{ij}) = (\omega_{jk} - \omega_{ij}) + (\omega_{ki} - \omega_{jk}) + (\omega_{ij} - \omega_{ki}) = 0.$$

Next, we derive a presentation of the operad $\mathbf{m}$. The degree 1 subspace of $\mathbf{m}(k)$ is one-dimensional for each $k \geq 2$, and is spanned by the operation

$$\{a_1, \ldots, a_k\} = \Delta(a_1 \ldots a_k) - \sum_{i=1}^{k}(-1)^{|a_1|+\cdots+|a_{i-1}|}a_1 \ldots (\Delta a_i) \ldots a_k.$$

This operation is clearly symmetric in $a_i$, and for $k=2$ is related to the Lie bracket in $\mathbf{bv}(2)$ by the formula $\{a_1, a_2\} = (-1)^{|a_1|}[a_1, a_2]$.

**Lemma 4.4.** *The operation $\{a_1, \ldots, a_k\}$ is given by the explicit formula*

$$\{a_1, \ldots, a_k\} = \sum_{1 \leq i < j \leq k}(-1)^{\varepsilon(i,j)}\{a_i, a_j\}a_1 \ldots \widehat{a_i} \ldots \widehat{a_j} \ldots a_k,$$

*where $\varepsilon(i,j) = (|a_1| + \cdots + |a_{i-1}|)|a_i| + (|a_1| + \cdots + |a_{j-1}|)|a_j| + |a_i||a_j|$.*

*Proof.* It is immediate from the definition of the bracket $[a, b]$ that

$$\{a_1, \ldots, a_k\} = \sum_{i=1}^{k-1}(-1)^{|a_1|+\cdots+|a_i|}a_1 \ldots a_{i-1}[a_i, a_{i+1} \ldots a_k] = 0.$$

Applying the Poisson rule, we see that this equals

$$\sum_{1 \leq i < j \leq k}(-1)^{|a_1|+\cdots+|a_i|+(|a_i|-1)(|a_{i+1}|+\cdots+|a_{j-1}|)}a_1 \ldots \widehat{a_i} \ldots [a_i, a_j] \ldots a_k.$$

A little manipulation of signs yields the lemma. $\square$

**Theorem 4.5.** *The operations $\{a_1, \ldots, a_k\}$ generate the gravity operad $\mathbf{m}$, and all relations among them follow from the (generalized) Jacobi relations*

$$\sum_{1 \leq i < j \leq k}(-1)^{\varepsilon(i,j)}\{\{a_i, a_j\}, a_1, \ldots, \widehat{a_i}, \ldots, \widehat{a_j}, \ldots, a_k, b_1, \ldots, b_\ell\} = \{\{a_1, \ldots, a_k\}, b_1, \ldots, b_\ell\},$$

*where the right-hand side is interpreted as zero if $\ell = 0$.*

*Proof.* Since the action of $\Delta$ on $\mathbf{b}(k)$ has vanishing cohomology, we may identify $\mathbf{m}(k)$ with the image of $\Delta : \mathbf{b}(k) \to \mathbf{b}(k)$. The generalized Jacobi relation follows by applying $\Delta$ to the equation in $\mathbf{b}(k+\ell)$

$$\{a_1,\ldots,a_k\}b_1\ldots b_\ell = \sum_{1\leq i<j\leq k} (-1)^{\varepsilon(i,j)}\{a_i,a_j\}a_1\ldots \widehat{a_i}\ldots \widehat{a_j}\ldots a_k b_1\ldots b_\ell,$$

which follows from Lemma 4.4.

The $k$-linear operations of degree $k-1$ on a gravity algebra are generated by the operation $\{a,b\}$, with the generalized Jacobi relation with $k=3$ and $\ell=0$ as the only relation:

$$\{\{a_1,a_2\},a_3\} + (-1)^{|a_1|(|a_2|+|a_3|)}\{\{a_2,a_3\},a_1\} + (-1)^{|a_3|(|a_1|+|a_2|)}\{\{a_3,a_1\},a_2\} = 0.$$

For this reason, we call such an operation a Lie monomial.

The free operad $\mathbb{T}\mathbf{b}(2)$ generated by $\mathbf{b}(2)$ consists of all words in a graded commutative product of degree zero, and a bracket of degree zero, with no conditions of associativity, Jacobi relations or Poisson relation imposed. We say that a word in $\mathbb{T}\mathbf{b}(2)$ is normalized if it is a product of Lie monomials. Since $\mathbf{b}$ is generated by $\mathbf{b}(2)$, there is a map of operads from $\mathbb{T}\mathbf{b}(2)$ to $\mathbf{b}$, obtained by imposing quadratic relations on $\mathbf{b}(2)$; this map is surjective when restricted to the normalized words. The result of applying $\Delta$ to a normalized word and taking its image in $\mathbf{b}$ is a word of the form $\{p_1,\ldots,p_\ell\}$, where $p_i$ are Lie monomials; this follows from the fact that $\Delta$ annihilates Lie monomials in $\mathbf{b}$. Thus, the operations $\{a_1,\ldots,a_k\}$ generate $\mathbf{m}$.

It remains to show that there are no further relations than the generalized Jacobi relations. Let $\mathbf{n}$ be the operad obtained by imposing on the generators of $\mathbf{m}$ all of the generalized Jacobi rules except those with $k > 3$ and $\ell = 0$. As vector spaces, $\mathbf{n}_i(k)$ and $\mathbf{b}_{i-1}(k)$ are isomorphic. The generalized Jacobi relations for $k > 3$ and $\ell = 0$ impose an additional $\dim(\mathbf{m}_{i-1}(k))$ relations on the $k$-linear expressions of degree $i$ in the operations $\{a_1,\ldots,a_k\}$; since $\dim(\mathbf{b}_i(k)) \cong \dim(\mathbf{m}_i(k)) + \dim(\mathbf{m}_{i-1}(k))$, this completes the proof of the theorem. $\square$

It remains to relate the operad $\mathbf{m}$ to topological string theory. If we follow the proof of [7] that the cohomology of a topological field theory $H_\bullet(\mathcal{E}, G_0^+)$ is a Batalin-Vilkovisky algebra, except that we substitute everywhere equivariant (co)homology for ordinary (co)homology, we obtain a proof of the following theorem.

**Theorem 4.6.** *Let $\mathcal{E}$ be the space of states of a two-dimensional topological field theory, with equivariant cohomology $H^\bullet_{S^1}(\mathcal{E})$. There is a natural structure of a gravity algebra on $H^\bullet_{S^1}(\mathcal{E})$.*

The physical significance of this result is subtle, since in most interesting cases, all but the lowest bracket vanish, and often even that one. However, we expect that gravity algebras will we useful in the study of Zwiebach's string field theory.

Department of Mathematics, MIT, Cambridge MA 02139 USA
*E-mail address*: getzler@math.mit.edu